# Restoring the privacy and confidentiality of users over Mobile collaborative learning (MCL) environment

[1]Abdul Razaque
[1]arazaque@bridgeport.edu

[2]Khaled Elleithy
[2]elleithy@bridgeport.edu

Wireless & Mobile communication laboratory
Computer science and Engineering department University of Bridgeport, CT, USA

*Abstract* — Mobile collaborative learning (MCL) is extensively recognized field all over the world. It demonstrates the cerebral approach combining the several technology to handle the problem of learning. MCL motivates the social and educational activities for creating healthier environment. To advance and promote the baseline of MCL, different frameworks have been introduced for MCL. From other side, there is no highly acceptable security mechanism has been implemented to protect MCL environment. This paper introduces the issue of rogue DHCP server, which highly affects the performance of users who communicate with each other through MCL. The rogue DHCP is unofficial server that gives incorrect IP address to user in order to use the legal sources of different servers by sniffing the traffic. The contribution focuses to maintain the privacy of users and enhances the confidentiality. The paper introduces multi-frame signature-cum anomaly-based intrusion detection systems (MSAIDS) supported with innovative algorithms, modification in the existing rules of IDS and use of mathematical model. The major objectives of research are to identify the malicious attacks created by rogue DHCP server. This innovative mechanism emphasizes the security for users in order to communicate freely. It also protects network from illegitimate involvement of intruders. Finally, the paper validates the mechanism using three types of simulations: testbed, discrete simulation in C++ and ns2. On the basis of findings, the efficiency of proposed mechanism has been compared with other well-known techniques.

I. INTRODUCTION

The mobile devices fill the gap of communication by providing the matured platform for learners. This extremely emerging platform has built the foundation to introduce the robust and faster way for promoting the pedagogical activities. The exploitation of mobile devices has not only encouraged the MCL but also created several chances for intruders attackers to break the privacy of users. The mobile users are reliant with DHCP server for getting IP address. The DHCP server offers well-structured and administrative service for mobile devices. From other side, illegal and Rogue DHCP server is installed in the network to capture the legal contents of users. Rogue DHCP server creates opportunity for intruders to intercept network traffic. After capturing the contents, intruder modifies the original contents of legal communicators. The intruder uses the malware and Trojans horse to install rogue DHCP server automatically on network.. If the rogue DHCP server assigns an incorrect IP address faster than original DHCP server, it causes the potentially black hole for users. To control the malicious attacks and avoiding the network blockage, the network administrators put their efforts to guarantee the components of server, using various tools. The graphical user interface (GUI) tool is used to prevent the attack of rogue detection [5]. Idea of using multilayer switches may be configured to control the attacks of rogue DHCP server but it is little bit complex and not efficient to detect rogue DHCP server and its malicious consequences. According to statement of Subhash Badri, the representative of DHCP Server team mentions in his online report that GUI tool cannot make difference between malicious DHCP servers and erroneously configured rogue [7]. The DHCP spoofing is another solution for detecting rogue DHCP server. However, if single segment is spoofed that can damage the whole network. Spooling method takes long time till attacker has enough time to capture the traffic and assign the wrong IP addresses [8]. Time-tested, DHCP Find Roadkil.net's, DHCP Sentry, Dhcploc.exe and DHCP-probe provide the solution to detect and defend rogue DHCP server malware [6]. All of these tools cannot detect the new malicious attacks. Intrusion detection systems (IDS) are also introduced to ensure the protection of systems and networks. However, IDS cannot detect the intrusion due to increase in size of networks. The Signature based detection does not have capacity to compare each packet with each signature in database [2]. Distributed Intrusion Detection System (DIDS) is another technique to support the mobile agents. This technique helps the system to sense the intrusion from incoming and outgoing traffics to detect the known attacks [1]. Ant colony optimization (ACO) based distributed intrusion detection system is introduced to detect intrusions in the distributed environments. It detects the visible activities of attackers and identifies the attack of false alarm rate [3]. Anomaly based intrusion detection are introduced to detect those attacks for which no signatures exist [4], [6], [10].Both signature based and anomaly based IDS have not been used to detect the problems of rogue DHCP server. This paper introduces the multi-frame signature-cum anomaly based intrusion detection system supported with novel algorithms, inclusion of new rules in IDS and mathematical model to detect the malicious attacks and increase the privacy and confidentiality of users in MCL environment.

## II. PROBLEM IDENTIFICATION

Mobile collaborative learning (MCL) is getting high popularity in several fields. Everyone wants secure and confidential collaboration but it is vigorously affected due to intervention of intruder and attacker. The security flaws in MCL communication invite the attackers to modify the contents. The deployment of rogue DHCP server encourages the attackers to collapse the security. The rogue DHCP server issues incorrect IP address to user and in consequences, the user acknowledges negatively to wrong DHCP server. The attacker sniffs all the traffic sent by users to other networks causes the network violation. It breaks the security policies as well as privacy of users. MSAIDS handles this security flaws with support of novel algorithms, addition of new rules in current IDS and mathematical model

## III. RELATED WORK

The modern technologies and its deployment in computer and mobile devices have not only created new opportunities for better services but from other perspective, privacy of the users is highly questionable. The network-intruder and virus contagion extremely affect the computer systems and its counterparts. They also alter the top confidential data. Handling these issues and restoring the security of systems, IDS are introduced to control malicious attackers.IDS are erroneous and not providing the persistent solution in its current shape. The first contribution in the field of intrusion detection was deliberated by J.P Anderson in [28]. The author introduced notion about the security of computer systems and related threats. Initially, he discovered three attacks that are misfeasors, external penetrations and internal penetrations. The classification of typical IDS is discussed in [17]. The focus of the contribution is about reviewing the agent-based IDS for mobile devices. They have stated the problems and strength of each category of classification and suggested the methods to improve the performance of mobile agent for IDS design. Four types of attacks are discussed in [21] for security of network. They have also simulated the behavior of these attacks by using simulation of ns2. A multi-ant colonies technique is proposed in [25] for clustering the data. It involves independent, parallel ant colonies and a queen ant agent. They state that each process for ant colony takes dissimilar forms of ants at moving speed. They have generated various clustering results by using ant-based clustering algorithm. The findings show that outlier's lowest strategy for choosing the recent data set has the better performance. The contribution covers the clustering-based approach.

The discussed work in [23], implements the genetic algorithm (GA) with IDS to prevent the network from the attack of intruders. The focus of technique is to use information theory to scrutinize the traffic and thus decrease the complexity. They have used linear structure rules to categorize the activities of network into abnormal and normal behaviors. The work done in [18] is about the framework of distributed Intrusion Detection System that supports mobile agents. The focus of work is to sense the both outside and inside network division. The mobile-agents control remote sniffer, data and known attacks. They have used data mining method for detection and data analysis. Dynamic Multi-Layer Signature based (DMSIDS) is proposed in [2]. It detects looming threats by using mobile agents. Authors have introduced small and well-organized multiple databases. The small signature-based databases are also updated at the same time regularly. Algorithm is presented for adaptive network intrusion detection (ANID). The base of algorithm is on naive decision tree and naive Bayesian classifier [19]. The algorithm performs detection and keeps the track of false positive at balanced level for various types of network attacks. It also handles some problems of data mining such as dealing with lost attribute values, controlling continuous attributes and lessening the noise in training data. Work is tested by using KDD99 benchmark for intrusion detection dataset. The experiment has reduced the false positive by using limited resources. Moreover, all of the proposed techniques cover general idea of network detection but proposed MSAIDS technique handles the irreplaceable issues of DHCP rogue server. It controls signature and anomaly based attacks to be generated by DHCP rogue. The contribution also prevents almost all types of attacks. The major contribution of work is to validate the technique by employing innovative algorithms, inclusion of new rules in tradition IDS and mathematical model. It also helps the legitimate users to start secure and reliable MCL frequently. One of the most promising aspects of this research is uniqueness because there is no single contribution is available in survey about the DHCP rogue and its severe targeted attacks. The reminder of paper is organized as follows: possible attacks of rogue DHCP server are explained in section-4. The proposed solutions including functional components are given in section-5. Simulation setup is explained in section-6. The analysis of result and discussion are given in section-7. Finally conclusion of the paper is given in section-8.

## IV. POSSIBLE ATTACKS OF ROGUE DHCP SERVER (SCENARIO-I)

With deployment of latest technologies, the need for automated tools has been increased to protect the information stored either on computers or flowing on networks. The generic idea to protect the data and thwart the malicious attackers is computer security. The introduction of distributed system has highly affected the security [11].
There are several forms of vulnerabilities and vigorous threats to expose the security of the systems. To take important security measures and enhancing the secure needs for organizations, several mechanisms are implemented but those mechanisms also invite the attackers to play with privacy and confidentiality of users. One of the major threats for privacy of data is intervention of rogue DHCP server.

The first sign of problem associated with rogue DHCP server is discontinuation of network service. The static and portable devices start experiencing due to network issues. The issues are started by assigning the wrong IP address to requested clients to initiate the session. The malicious attackers take the advantages of rogue DHCP server and sniff the traffic sent by legitimate users.

Rogue DHCP server spreads the wrong network parameters that create the bridge for attackers to expose the confidentiality and privacy. Trojans like DNS-changing installs the rogue DHCP server and pollutes the network. It provides the chance to attackers to use the compromised resources on the network. Rogue DHCP server creates several problems to expose the privacy of legitimate users. The most three important attacks are shown in figure 1.

legitimate DHCP server. The rogue DHCP server releases the wrong network settings and in consequence causes the network disruption.

The DHCP server responds faster than corporate server on local area network for release of IP addresses. Rogue DHCP server continuously monitors the behavior of DHCP server and whenever any request comes through router for release of IP address, it issues the wrong IP address faster than legitimate DHCP server. Router does not have capability to detect issued IP address from rogue DHCP server. Once it issues its wrong IP address then continuously disrupts the network. Several mechanisms are introduced such as network access control (NAC) and network access protection (NAP) to force all the connected devices to be responsible for authentication to network.

The devices which do not meet the criteria of authentication are segmented on subnet or virtual Local area network (VLAN). All of these methods are challenging and difficult to deploy to protect the network disruption from attack of rogue DHCP server. To employ these mechanism, DHCP snooping must be implemented on switches otherwise users will not be able to obtain the information but from other side NAC and NAP are limited with some specific operating systems. These techniques are not efficient to control the attack rogue DHCP server.

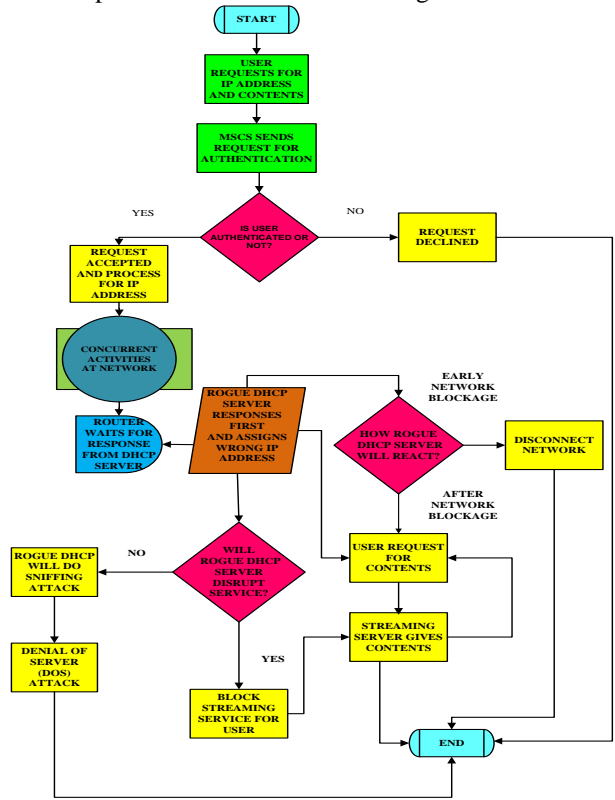

Figure 1. Behavior of DHCP Rogue during the attack

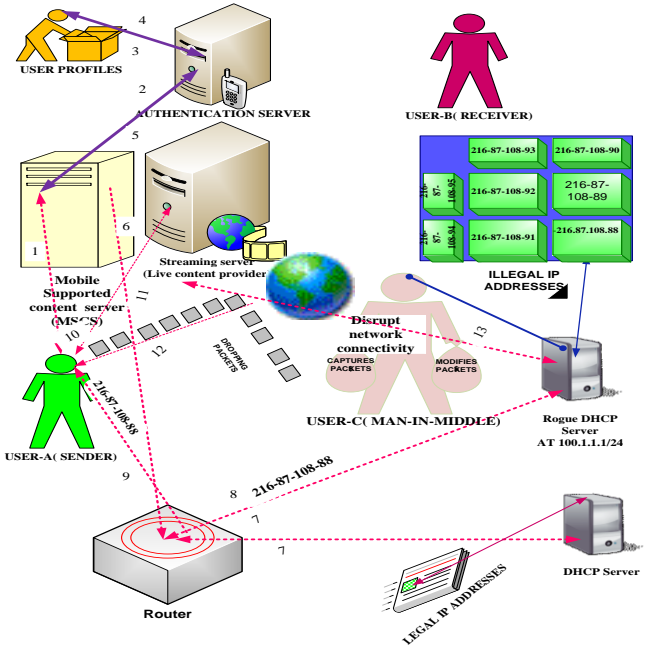

Figure. 2. Network disrupting attacking process

### A. Network disrupting attack

Rogue DHCP server makes the vague situation for legitimate DHCP server by sending many requests for issuance of IPs. This situation exhausts the pool of IP addresses of DHCP server, resulting it does not entertain the requests of clients for lease of new IPs. When process of issuance of IPs is blocked then rogue DHCP server appears in network as replacement of

It takes the control of whole network and also continues the connecting and disconnecting process at different intervals. The figure 2 shows the attacking process of DHCP rogue server and disruption of network.

## B. Sniffing the network traffic

It is brutal irony in information security that the features which are used to protect the static and portable devices to function in efficient and smooth manner; and from other side same features maximize the chances for attackers to compromise and exploit the same tools and networks. Hence packet sniffing is used to monitor network traffic to prevent the network from bottleneck and make an efficient data transmission. Attackers use the same resources for collecting information for illegal use. Rogue DHCP server substantiates those malicious attackers to expose the privacy of users. When networks are victim of rogue DHCP servers that provide very important information related to IP address, domain name system and default gateway to attackers.

All of this information helps the intruders to sniff the traffic of legitimate users. Suppose rogue DHCP server is introduced on secure environment to collect the confidential information. That server destroys all sorts of havoc on secure network. In the best case scenario, it simply issues the wrong IP address to each user, resulting all the traffic on the network starts monitoring on the bases of issued wrong IP address. In the worst case scenario, rogue DHCP server sets default gateway as IP address of malicious attacker's proxy. In this case, attacker can sniff the traffic and wreak the privacy of users shown in figure 3.

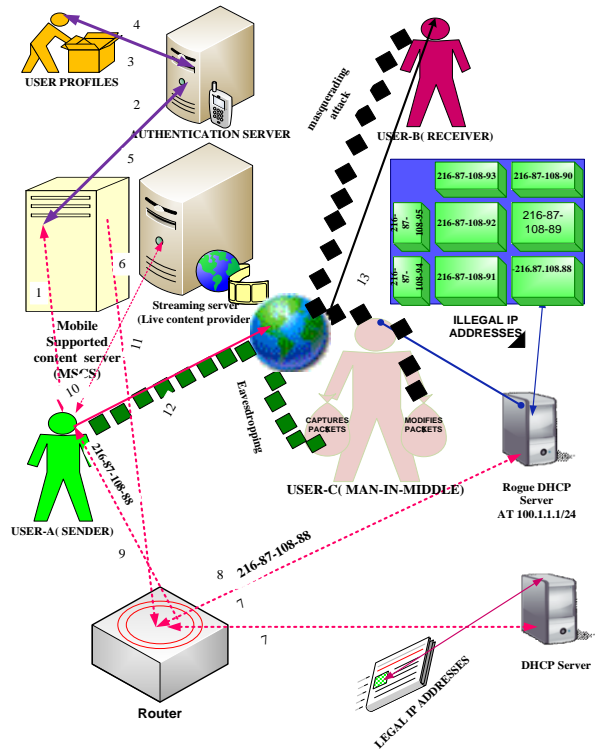

Figure 3. Sniffing the traffic and masquerading attack

Rogue DHCP server also helps the intruders to capture the MAC address of legitimate users. It causes the sniffing the traffic through switch. In this case, attackers spoof the IP addresses of both sender and receiver and play the man-of-middle to sniff the traffic and extract important contents of communication. It causes the great attack on privacy of users. The figure 4 shows the process how the IP of both sender and receiver are spoofed and privacy is exposed. This type of attack refers as masquerading attack.

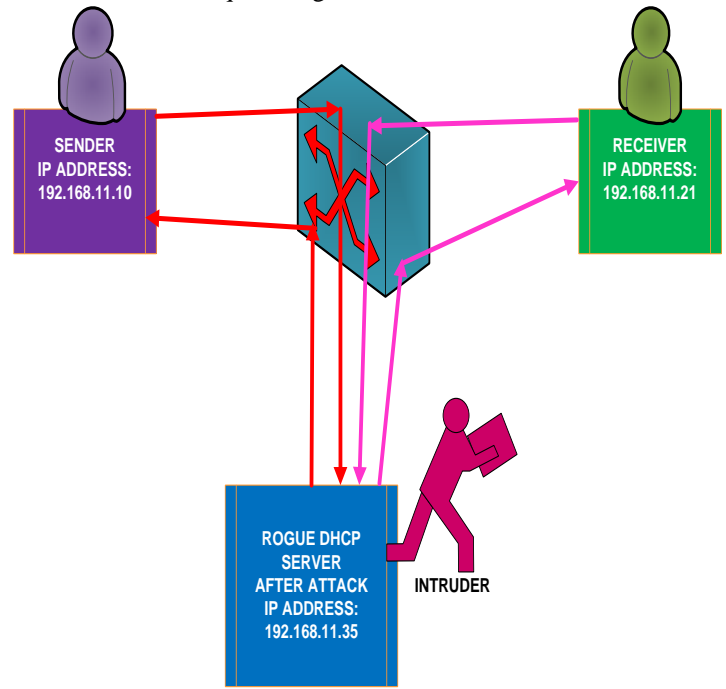

Figure 4. Spoofing of IP address and diversion of traffic

## C. Denial of service attack ( DOS)

Intruders that get support through rogue DHCP server also use DOS attacks after sniffing the confidential contents of traffic. Due to DOS attack, the access of important services for legitimate users is blocked. Intruders often crash the routers, host, servers and other computer entity by sending overwhelming amount of traffic on the network. Rogue DHCP server creates friendly environment for intruders to launch DOS attacks because intruders need small effort for this kind of attack and it is also difficult to detect and attack back to intruders. In addition, it is also easy to create the floods on internet because it is comprised of limited resources including processing power, bandwidth and storage capabilities. Rogue DHCP can make flooding attack at the domain name system (DNS) because target of intruders is to prevent the legitimate users from resolving resource pertaining to zone under attack [12] [13] & [16].

These attacks on DNS have obtained varying success while disturbing resolution of names related to the targeted zone. Rogue DHCP server can take advantages of inevitable human errors during installation, configuration and developing the software. It creates several types of DOS attack documented in literature [20]. [22] & [24]. Intruders with support of Rogue DHCP server make three types of fragile or smurf, SYN Flood and DNS Attacks DOS attacks shown in figure 5.

These attacks are vulnerable and dangerous for security point of view.

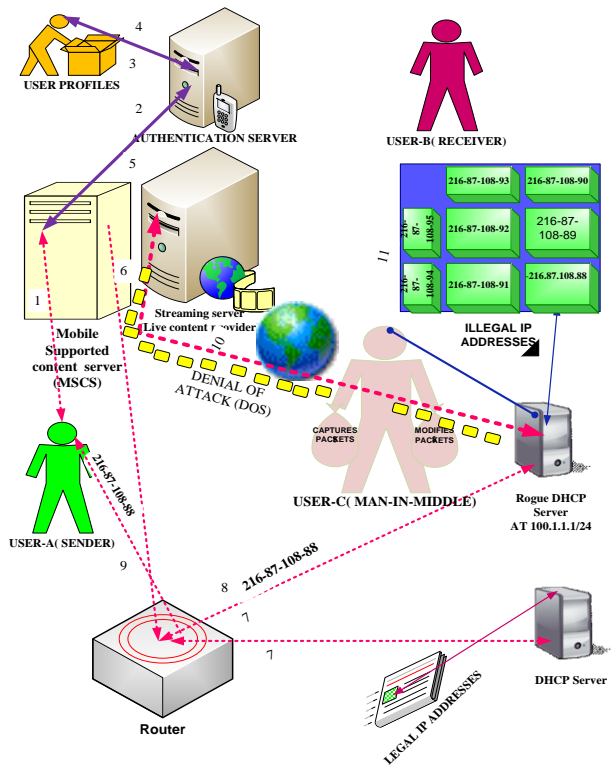

Fig.ure 5.denial of service attack (DOS) attack

## V. PROPOSED SOLUTION(MULTI- FRAME SIGNATURE-CUM-ANOMALY BASED INTRUSION DETECTION SYSTEM)

Networks are being converged rapidly and thousands of heterogeneous devices are connected. The devices integrated in large networks, communicate through several types of protocols and technologies. This large scale heterogeneous environment invites the intruders to expose the security of users. Hence, IDS are introduced to recognize the patterns of attacks, if they are not fixed strategically, many attackers cross the IDS by traversing alternate route in network.

Many signature-based IDS are available to detect the attacks but some of new attacks cannot be identified and controlled. Anomaly-based IDS is another option but it can only detect new patterns of attack. The multi-frame signature-cum anomaly-based intrusion detection system (MSAIDS) supported with algorithms is proposed to resolve the issue of DHCP rogue. The proposed framework consists of detecting server that controls IDS and its related three units: (i) DHCP verifier unit (ii) signature database and (iii) anomaly database.

During each detection process, intrusion detection starts matching from DHCP verifier, if any malicious activity is detected that stops the process otherwise checks other two units until finds and malicious activity or not. Figure 6 shows the MSAIDS.

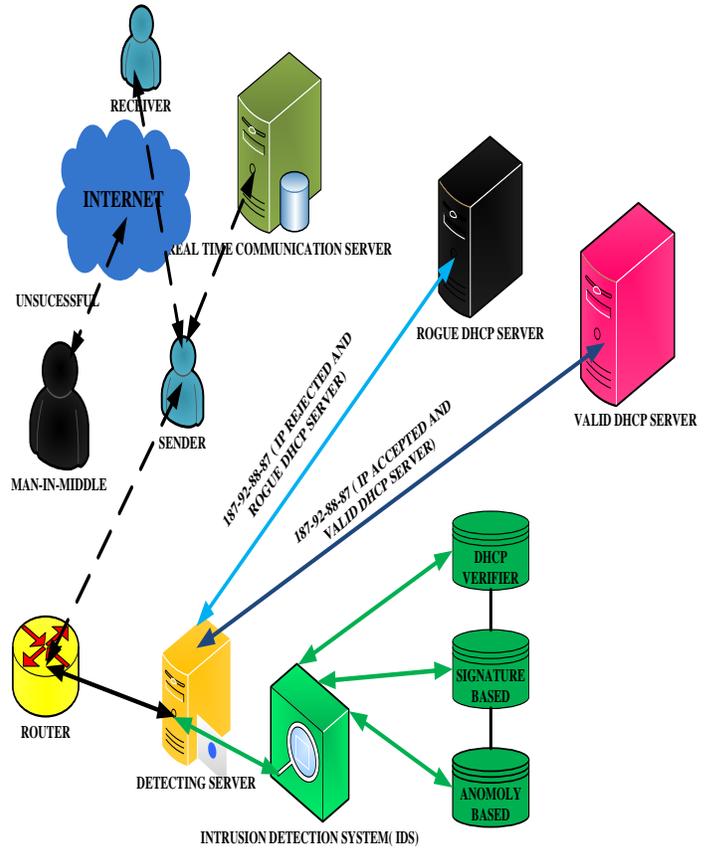

Figure 6. Multi-frame signature-cum-anomaly based IDS

The detecting server (DS) is responsible to check the inbound and outbound traffic for issuance of IP address. The DS gets IP request (inbound traffic) from routers and forwards to DHCP server after satisfactory checkup. When any IP address is released for requested node then applies DHCP detecting algorithm for validation of DHCP server and detecting the types of attack shown in algorithm 1.

**Algorithm 1: Verify DHCP server and detecting the attack**

1. Input: MF =(FD, FS,FA & I)
2. Output : For every strategy I € FA, I € FS, D € FD)
3. D = Each valid DHCP Server
4. IP= Internet protocol address
5. N= Number of mobile devices
6. FD= Frame DHCP server
7. If D € FD
8. IP→ N
9. endif
10. S= Number of available signatures in signature based Intrusion detection system (SIDS)
11. FS= Frame of signatures
12. FS ⊆ SIDS
13. I= Number & Types of attacks
14. For ( I=S;  I ≤ FS; I++)
15. If  I ⊆ FS
16. SIDS attack alert
17. endif

18. endfor
19. A= Number of signatures available in Anomaly based Intrusion detection system AIDS
20. FA= Frame of AIDS
21. FA ⊆ AIDS
22. For ( I =A; I ≤ FA ; I ++)
23.  If  I  ⊆ FA
24. AIDS raises alert
25. If ( I ∉ FS & I ∉ FA)
26.  No alert ( No attack)
27. endif
28. endif
29. endfor

A. *The monitoring process of detecting system DS*

 i. Pre-selected rule refers to those patterns, which are already stored in DS that helps to identify the inbound traffic.

 ii. Post-selected rule refers to those patters which are stored for detection of legitimate DHCP server that helps to identify outbound traffic.

 iii. Parameterized rule refers to many ingredients that help to set selected rules with unique value presented in the following:

   a. **Validity ingredient:** It helps to detect the attack if intruder modifies the contents of message.

   b. **Time interval ingredient:** It helps to detect two types of attacks which are exhaustion attack and negligence attack. In exhaustion attack, the attacker increases message-sending rate. In negligence attack, attacker does not send the message. In addition, time interval for two consecutive messages is extended or lessened than allowed amount of time, attack is considered.

   c. **Flooding ingredient:** It helps to identify the attack on the basis of noise and disturbance to be created in the communication channel.

   d. **Retransmission ingredient**: It helps to determine the attack, if retransmission does not occur before specified timeout period.

   e. **High transmission radio range ingredients**: It helps to determine SYN flood and wormhole attack, when intruder uses powerful radio sending a message to further located node.

   f. **Pattern replication ingredient:** It helps to detect the attack when same patterns are repeated several time, it blocks the DOS attacks.

All of these ingredients collectively help to DS to detect the attacks and figure 7 shows the process how to determine valid IP address and attack.

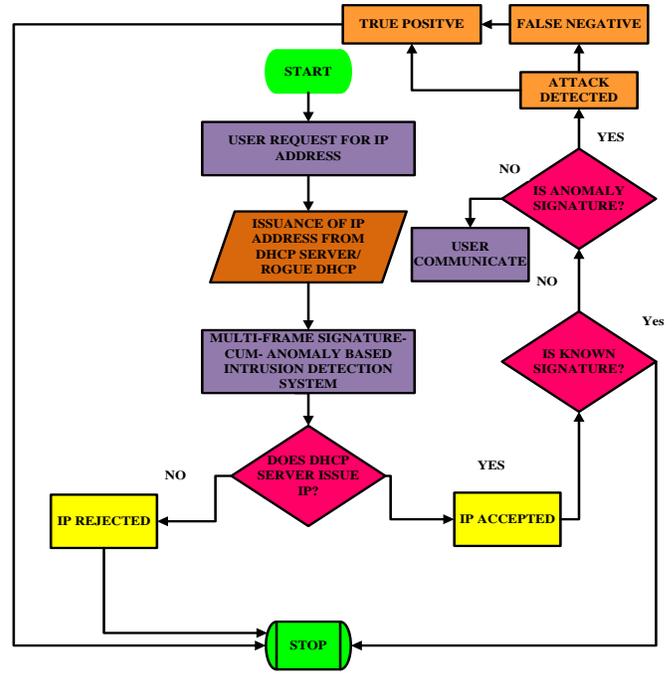

Figure 7. Detecting attack and issuance of safe IP

 DS also controls the multi frame that comprises of central IDS and integrated with three layers that control the misuse detection.

B. *Central IDS*

The aim of central IDS is to control and store messages received from DS. It works as middleware for DS and other layers to send the verification request and receive the alerts. The main function of central IDS is to update and manage the policy according to attack. If it needs any change in attack detection that is employed on all the layers. The central IDS implements the updated policy is shown in figure no 8.

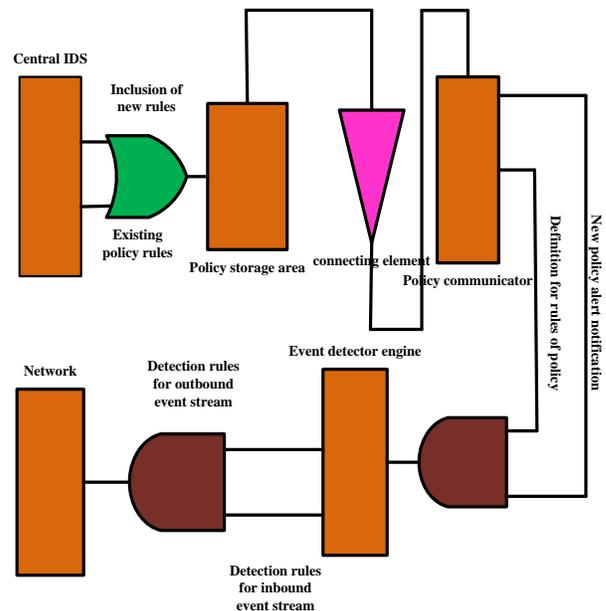

Figure 8. Policy of Central IDS for network

## C. DHCP Verifier

DHCP verifier is the top layer that distinguishes between rogue DHCP and original DHCP server. The signatures of original DHCP servers are stored at the DHCP verifier. It checks the validity of the DHCP server which issues the IP address for client. On the basis of stored signatures, DHCP server is identified whether it is rogue or original DHCP server. Top layer produces the unique alert sign for both DHCP rogue and original DHCP. Top layer receives the parameters for verification from central IDS. DHCP verifier running on top layer is also responsible to return the alert to central IDS.

## D. Signature based detection layer

Signature-based detection is middle layer that detects known threat. It compares the signatures with observed events to determine the possible attacks. Some known attacks are identified on the basis of implemented security policy. For example, if telnet tries to use "root' username that is violating the security policy of organization that is considered the known attack. If operating system has 645 status code values that is indication of host's disabled auditing and refers as attack.

If attachment is with file name "freepics.exe" that is alert of malware. Middle layer is effective for detection of known threats and using well-defined signature patterns of attack. The stored patterns are encoded in advance to match with network traffics to detect the attack. This layer compares log entry with list of signatures by deploying string comparison operation. If signature based layer does not detect any network, anomaly based detection layer initiates the process. For example, if request is made for web page and to be received the message with status code of 403, it shows that request is declined and such types of processing cannot be tailored with signatures based layer. The figure 9 shows the combination of IDS rules applied to determine the various types of attacks with support of *MSAIDS*.

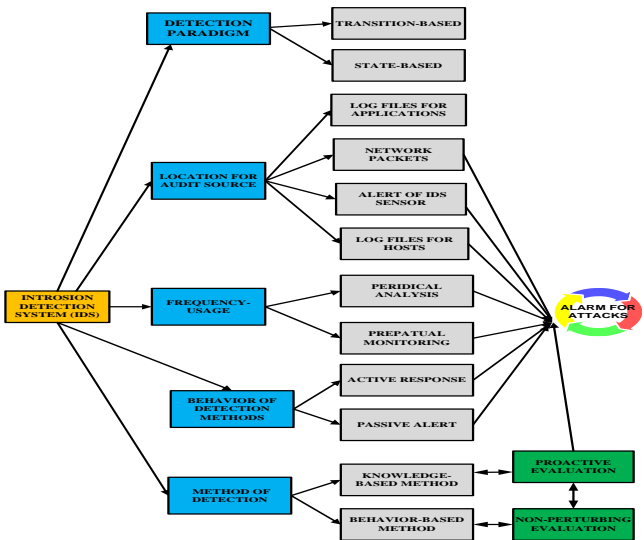

Figure 9. IDS rules for detection the attacks

The base of IDS comprises of five fundamental rules which support to IDS to determine the attack. These rules include detection paradigm, location for audit source, frequency usage, method for detection and behavior of detection method.
All of these rules help to detect the type of the attacks which are created due rogue DHCP server.

## E. anomaly based detection layer

Lower layer is anomaly based detection that identifies unknown and DOS attacks. It works on pick-detect method. This method monitors the inbound and outbound traffic received through central IDS. Packets are evaluated, adaptive thresholds and mean values are set. It calculates the metrics and compares with thresholds [26] & [27].

On the basis of comparisons, it detects various types of anomalies including false positive, false negative, true positive and true negative. If pick-detect methods determines true positive and false negative then it sends alert to Central IDS. The process of determining the anomalies is given in algorithm 2.

**Algorithm 2: Detecting the types of alerts with AIDS**

1. FS= Frame of signatures
2. FS ⊆ SIDS
3. Si = False Negative
4. Sj= True negative
5. Sk= True positive
6. Sk= False positive
7. 0 = don't match & 1= match
   m
8. Sijkl = 1/d ∑  Sijkl
               m=1
9. Sij = { 0, if i & j
10. No false negative & true positive
11. Sij = { 1, if i & j
12. false negative & true positive
13. Alert of attack
14. Skl = { 0, if k & l [ do not match] & 1, if k & l [match]
15. Alert of true negative & false positive
16. No sign of attack
17. endif
18. endif
19. endif
20. endif

In addition to determine and calculate true positive, true negative, false positive and false negative, the following derivation helps:
Here True positive= TP; False negative= FN; False positive=FP; True negative=TN; Precision= p; overall probability= OP
We know that
P = TP/ TP + FP ----------------------------------(1) Eq:
OP= TP + TN / TP + FP + FN + TN ----------- (2) Eq:
TP = P (TP +FN) ---------------------------------(3) Eq:
Substitute the values of precision in equation no (2)

Therefore,
 TP= TP/ TP + FP (TP +FN) -------------------- (4) Eq:
OR
TP = (TP * TP) + (TP + FN) / TP + FP -------- (5) Eq:
OR

**TP= TP$^{2+}$ (TP* FN) / TP + FP**

The equation (5) shows the true positive (TP), meaning that is sign of attack and alarm occurs

Now, find the false negative to apply overall probability formula:

We know that overall probability formula that is given as follows:
OP= TP + TN / TP + FP + FN + TN
TP = OP (TP + FP + FN + TN) = TP + TN ----------- (6) Eq:
Substitute the value of OP in equation (6):
TP = TP + TN / TP + FP + FN + TN (TP + FP + FN + TN) = TP + TN-------------------------------------------------- (7) Eq:
 (TP + TN) * TP + (TP + TN)* FP + (TP + TN) *FN + (TP + TN)*TN / TP + FP + FN + TN =
TP +TN-------------------------------------------------- (8) Eq:
Multiplying the values:
(TP$^{2+}$ TNTP) + (TP*FP + TN*FP) + (TP*FN + TN*FN) + (TP*TN + TN$^2$) = TP + TN (TP + FP + FN + TN) --- (9) Eq:
Re-arranging:
(TP$^{2+}$ TNTP) + (TP*FP + TN*FP) + (TP*TN + TN$^2$)=( TP + TN ( TP + FP + FN + TN) - (TP*FN + TN*FN)
Multiplying the (TP + TN) to the right hand side:
(TP$^{2+}$ TN*TP) + (TP*FP + TN*FP) + (TP*TN + TN$^2$) = (TP$^2$ + TN*TP +FP*TP +FP*TN+ FN*TP + FN*TN + TN*TP + TN$^2$ - (TP*FN +
TN*FN) --------------------------------------------------- (10) Eq:
OR
TP$^2$ + TN*TP + TP*FP + TN*FP + TP*TN + TN$^2$= TP$^2$ + TN*TP +FP*TP +FP*TN+ FN*TP + FN*TN + TN*TP + TN$^2$ - TP*FN - TN*FN
Re-arranging the terms:
TP$^2$ + TN*TP + TP*FP + TN*FP + TP*TN + TN$^2$ - TP$^2$ - TN*TP -FP*TP - FP*TN- FN*TP - FN*TN - TN*TP - TN$^2$ + TP*FN + FN*TN
Simplifying the terms by using addition and subtraction function:
FN*TN - FN*TN --------------------------------------- (11) Eq:
Finding the value of FN
FN = FN * TN /TN ------------------------------------ (12) Eq:
Divide TN with FN *TN to get FN
FN= FN --------------------------------------------------- (13) Eq
OR

**FN-FN= 0**

If we get zero value that shows the false negative and considered as attack but no sign of alarm because of 0.

The false positive (FP) is derived from true positive (TP). If We know that about the value of TP:
TP= TP$^2$ + (TP * FN) / TP + FP
Cross by multiplication to both sides:
TP (TP + FP) = TP$^{2+}$ (TP + FN) ------------------------ (14) Eq:
Multiplying TP to the left hand side:

TP$^2$ +TP*FP = TP$^{2+}$ (TP * FN) -------------------------- (15) Eq
Re-arrange the terms:
FNTP =TP$^{2+}$ TP * FP - TP$^2$
FN*TP = TP * FP ----------------------------------------- (16) Eq:
Calculate the FP from equation (16):
FN*TP= FP/TP ----------------------------------------- (17) Eq:
OR

**FP= FNTP /TP**

 The equation (17) shows the false positive; there is no sign of attack but alarm raised.

 Applying equation (10) to find the true negative (TN).
Here,
FN*TN - FN*TN
FN*TN = FN*TN
TN = FN*TN/ FN ------------------------------ (18) Eq:
Dividing FN*TN by FN to find TN
TN= TN ----------------------------------------- (19) Eq:
OR

**TN-TN=0**

The value of TN is also zero that means there is no sign of alarm and no attack occurs.

To detect the attack and non-attack situation for TN and FN. We use algorithm 3 to determine sign of attack.

**Algorithm 3: Determine the sing of attack or non-sign of attack**

I. We select random odd prime number for TN and any even number for FN.
2. The value of FN must not be exceeded than TN.
3. Therefore, FN > 1 & FN< TN
4. Here, FN= {2, 4, 6, 8…} & TN= {3, 5, 7, 11, 13…}
5. Here sign of attack = ST, d = not exposed & b = exposed.
6. b and d has constant value 1.
7. Thus, ST = TN/ (TN+ d)/ FN (FN +b)
8. If value of ST > 1, it means there is no sign of attack, if the value of ST < 1 that is sign of attack.
9. endif
Assume FN = 2 & TN =3
BY applying the sign of attack formula:
ST = TN/ (TN+ d)/ FN (FN +b)
Substitute the values in given formula.
ST = 3/ (3+ 1) / 2(2+1)
ST = 9/8
ST= 1.125
ST > 1

Here, ST > 1 means there is no sign of attack and we will be able to determine that is True negative (TN).

The behavior of TN, FN, FP& TP is shown in table 1 with respect to sign.

TABLE 1: showing the behavior of attack

| parameters | Behavior |
|---|---|
| True Negative (TN) | No sign of alarm and no attack |
| False Negative (FN) | No sign of alarm but attack |
| True positive ( TP) | Sign of alarm and attack detected |

| False positive (FP) | Sign of alarm but no attack |
| 2.4/5GHz | |

## VI. SIMULATION SETUP

The previous sections have presented the evidence of the problems to be created by rogue DHCP server and including solutions to control over these problems. This section focuses on simulation setup and type of scenario. The unique and extensive testing method require for simulation of rogue DHCP server. To validate the framework, the proposed solution has been implemented by using three methods: test bed simulation, discrete simulation in C++ and ns2 simulation. The test bed simulation provides real time results in controlled and live user environments. This kind of simulation gives complete understanding about behavior of several types of attacks. All operations associated with MSAIDS approach and other three existing approaches: Dynamic Multi-Layer Signature based IDS (DMSIDS), Ant Colony Optimization based IDS (ACOIDS) & Signature based IDS provide the recital idea. Test bed simulation helps to detect the various affects and conditions in real environment. Therefore, output of simulated framework is as close to reality as possible. The parameters of test bed simulation are only given in table 2.

TABLE 2: Simulation parameters for test bed experiment

| Name of parameters | Specification |
|---|---|
| MySQL database | MySQL 5.5 |
| Type of IDS | Rule based IDS |
| GD Library | gd 2.0.28 |
| Snort | V-2 |
| Apache web server | Apache http 2.0.64 Released |
| PHP | PHP 5.3.8 (General purpose server side language) |
| ADODB | Release 5.12 (abstraction library for PHP and Python) |
| ACID | ACID PRO 7 |
| Stick | Stick beats detection tool used by hackers |
| Nikito | Nikito v.2.1.4 |
| IDS enabled system | Memory: 512 MB |
| | Operating system: Linux |
| | PCI network card: 10/100 Mbps |
| | CPU: P-III with 600 MHz |
| Attacker system | Memory: 1.5 GB |
| | Operating system: Linux |
| | PCI network card: 10/100/1000 Mbps |
| | CPU: AMD Geode LX running |

In addition, stressed network traffic load creates several exceptional conditions that motivate to test the idea on Linux and window NT operating systems. It is little bit harder to collect the data in live environments by using test bed simulation due to some logistical issues. Most of operating systems do not provide the tracing facilities but regardless of problems, we would like to obtain the result in standardized method by using different programs on different operating systems. MSAIDS has fully support of algorithms and data structure that discover the potential attacks and perturb the intrusions before the attacks.

The performance highly depends on robust tracing facility and algorithms, which help to identify the intrusion. As the first step is to analyze the performance of proposed algorithms and used mathematical model. However, overall target is to obtain accurate statistical data in highly loaded network. Test bed simulation does not provide 100% result. To fulfill requirements of tracing and collecting accurate data, same scenario is simulated in ns2 and discrete simulation c++. The mean value is calculated for three kinds of simulations with help of following theorem.

**Theorem 1:**

Assume x = test bed simulation;
y= discrete simulation in C++ & ns2 simulation.
R is the proposed approach MASIDS.
Thus, Let f: [x, y] → R is the continuous function for closed interval [x, y]
Therefore
Let f: [x, y] → R is the differentiable continuous function for open interval (x, y)
Here x < y.
Hence z exists in (x, y)
Such that

$$f'(x) = \frac{f(x) - f(y)}{x - y}$$

The second important step is in context of IDS what is the most important response, if once possible attack has been detected? It is really broader topic and beyond of the scope of this research. The proposed method is not panacea but at least detects masquerading, DOS and race condition attacks to be generated by DHCP rogue in MCL. The idea of controlling the attack of rogue DHCP server is quite general but might be efficient to enhance the security of asynchronous teller machine (ATM) machines, heterogeneous network environment, COBRA and distributed object systems.

The paper emphasizes how to fit proposed method to control the attacks of rogue DHCP server in MCL environment. Figure 10 is NAM screenshot of ns2 for MSAIDS, in which node-3 is rogue DHCP server that assigns the fake IP. Node-4 is attacker that tries to capture the traffics. Node-5 is MSAIDS (proposed approach) that makes an attack of node-4 (attacker) in vain. In consequences, sender and receiver exchange the data successfully.

The scenario is highly congested with large amount of traffic in real but standardized environment.

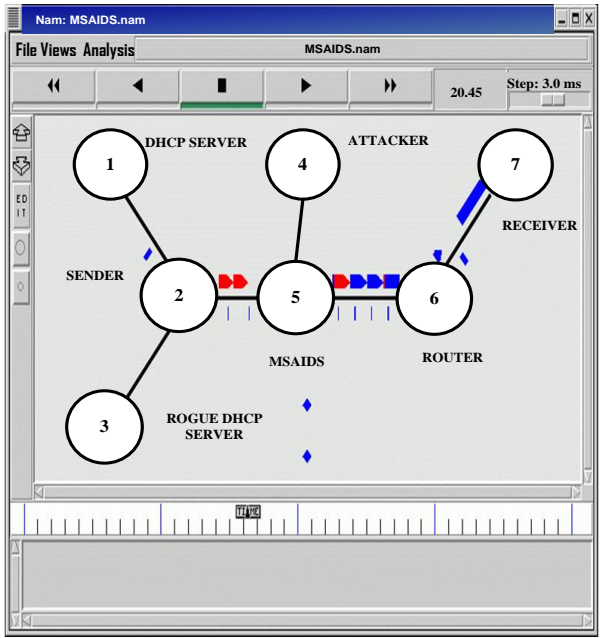

Figure.10. NAM screenshot of a MSAIDS

## VII. ANALYSIS OF RESULT AND DISCUSSION

The training period of experiment covers four classes of attacks probe, DOS, U2R and R2L. All detected attacks are included in database during the training period in test bed simulation. Ns2 and discrete simulation in C++ provides tracing facility to collect accurate data. The MSAIDS scans all rules of snort and includes new rules explained in proposed section 3. The testing period targets one concise scenario. The scenario is simulated by using same parameters for all three existing approaches including our proposed MSAIDS. The attacks are generated by using stick, covering all types of signatures and anomaly based attacks.

The training period provides quite interesting results because frequently generated attacks are of different numbers. The maximum number of attacks pertains to R2L category. The more attacks are also counted on MSAIDS as compare with other three existing techniques are shown in table no.3. If attack is not generated then it is counted as normal traffic. The frequency of single and group characters is displayed when packets reach at the attacker machine. It is observed on the basis of output that different types of detected attacks are generated due to rogue DHCP server.

The DOS attacks are detected when packet does not reach at destination and received no acknowledgment. The sign of probe attack is addition of new data in existing amount of data bytes. U2R is the sign of maximum connection duration. R2L attacks are little bit complex to detect. We apply method comprises of service requested and duration of connection for network and attempts failed login for host. It shows that proposed approach does not restrict the generating ratio of packets. From other side, the proposed approaches provides highest capturing ratio. The statistical results show that MSAIDS will substantiate to medical field for diagnosing several disease and especially for heart. The major breakthrough of this research is to detect the true positive and false negative attacks because they are very hard to capture.

TABLE 3: showing statistical data for attacks

| Type of generated attack | (MSAIDS) | (DMSIDS) | (ACOIDS) | Signature based IDS |
|---|---|---|---|---|
| DOS attacks | 34214 | 33542 | 33421 | 32741 |
| U2R attacks | 12454 | 11874 | 11845 | 11341 |
| R2L attacks | 34123 | 32123 | 31092 | 29984 |
| Probe attacks | 6214 | 8758 | 10181 | 4907 |

Due to these anomalies, confidentiality of any system is exploited and privacy of the user is exposed. The proposed method also captures the real worm attacks and all other looming attacks. The table 4 shows quantitative results for each scheme.

TABLE 4: Showing quantitative data for proposed and existing schemes

| Parameters | (MSAIDS) | (DMSIDS) | (ACOIDS) | Signature based IDS |
|---|---|---|---|---|
| total no: of packet to be received | 236719 | 198678 | 201938 | 178109 |
| total no: of packet to be analyzed | 236456 | 192453 | 197842 | 170098 |
| total no: of attack to be generated | 87005 | 86293 | 86539 | 78973 |
| total no: of signature based attack to be generated | 42003 | 42287 | 42585 | 35098 |
| total no: of anomaly based attack to be generated<br>a. False positive<br>b. False negative<br>c. True positive<br>d. True negative | a. 9475<br>b. 12093<br>c. 18574<br>d. 4860 | a. 8574<br>b. 11987<br>c. 16943<br>d. 6502 | a. 8878<br>b. 12007<br>c. 16943<br>d. 6126 | a. 8458<br>b. 11289<br>c. 12455<br>d. 11673 |
| Total number of anomaly based attack to be generated | 45002 | 44006 | 43954 | 43875 |
| total no: of attacks to be captured | 87002 | 84212 | 83434 | 36098 |
| Anomaly based attacked captured<br>a. False positive<br>b. False negative<br>c. True positive<br>d. True negative | a. 9475<br>b. 12093<br>c. 18574<br>d. 4859 | a. 8324<br>b. 11456<br>c. 15678<br>d. 6324 | a. 8678<br>b. 11987<br>c. 16789<br>d. 6045 | a. 654<br>b. 245<br>c. 453<br>d. 535 |
| total number of anomaly based attacks to be captured | 45001 | 41782 | 43499 | 1887 |
| signature based attacks to be captured | 42001 | 42102 | 41091 | 34211 |

The major advantage of MSAIDS approach is to detect all types of anomalies and unknown threats efficiently. The systems are mostly infected due to new sort of malwares because they consume the processing resources of system. If resources of system are utilized by unnecessary programs then MCL is highly affected. In consequence, collaboration process is disrupted.

MSAIDS also detects the activity for any specific session. It creates specific alarm for each type of anomalies. Furthermore, deployed algorithms and new addition of rules in ordinary IDS improves the performance and restore the privacy of users. The implementation of MSAIDS is supported with sound architectural design that is robust and can persist when attack is detected. Statistical data shows 99.996% overall efficiency of MSAIDS shown in table 5.

The efficiency of MSAIDS is calculated with following formula:

Here, overall efficiency = $E_a$;
Total generated signature based attacks = TSA;
Total anomaly based attacks =TAS;
Missed signature based attacks = MSA;
Missed anomaly based attacks = MAA & total generated attacks = TGA.

Thus, $E_a = (TSA + TAA) - MSA + MAA) * 100 / TGA$

TABLE 5: Overall efficiency of MSAIDS VS other approaches

| Parameters | MSAIDS | DMSIDS | ACOIDS | Signature based IDS |
|---|---|---|---|---|
| Packet analysis capacity (Mean value %) | 99.88 | 96.866 | 97.971 | 95.5 |
| Anomaly based attacked captured (Mean value %) a. False positive b. False negative c. True positive d. True negative | a. 100 b. 100 c. 100 d. 99.979 | a. 97.084 b. 95.570 c. 92.533 d. 97.262 | a. 97.747 b. 99.833 c. 99.091 d. 98.677 | a. 7.732 b. 2.170 c. 3.637 d. 4.583 |
| Total number of anomaly based attacks to be captured (Mean value %) | 99.997 | 94.946 | 98.964 | 4.3 |
| signature based attacks to be captured (Mean value %) | 99.995 | 99.562 | 96.491 | 97.472 |
| Overall efficiency of different approaches | 99.996 | 97.254 | 97.727 | 50.886 |

The proposed MSAIDS framework produces 2.269 to 49.11 higher capturing-rates than other existing techniques. The more interesting work of this research is detailed expression of all types of anomalies separately in form of false positive, false negative, true positive and true negative. These parameters give the concrete idea to use the features for various types of applications in real environment. The figure 10 shows the capturing capacity of MSAIDS and other existing techniques with respect the time during the testing period.

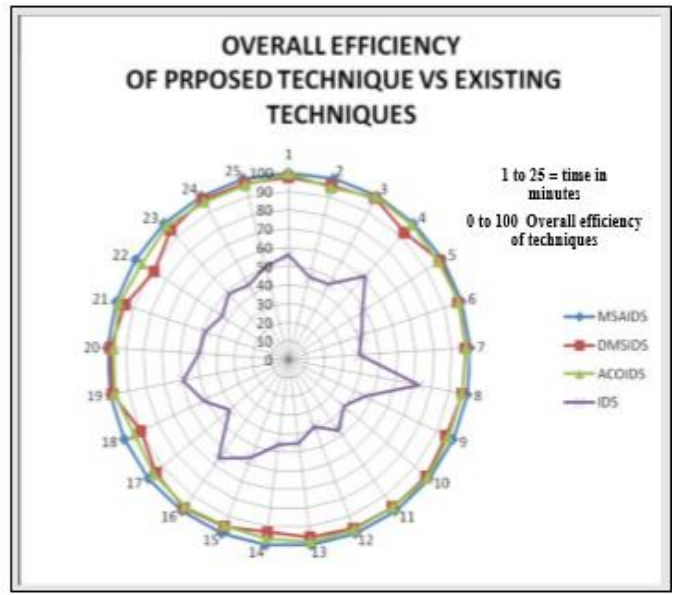

Figure 11. Comparison efficiency of all approaches

## VIII. CONCLUSION

In this paper, multi-frame signature-cum anomaly-based intrusion detection systems (MSAIDS) is presented. MSAIDS controls malicious activities of DHCP rogue server to restore the privacy of users during MCL. The paper highlights all the malicious threats to be generated by DHCP rogue. The intruders use the DHCP server to sniff the traffic and finally deteriorate the confidential information. The mechanism of current IDS does not have enough capability to control the threats. Furthermore, several daunting and thrilling challenges in the arena of computer network security are impediment for secure communication. DHCP rogue is visibly very simple but crashes the network as well as the privacy of the users and even creates nastier attacks like Sniffing network traffic, masquerading attack, shutting down the systems and DOS. The first is detailed explanation of these attacks and how they are generated by DHCP rogue.

To resolve this issue, second propose the technique that is based on algorithms, mathematical modeling and addition of new rules in current IDS. These all of the components of proposal collectively handles the issues of DHCP rogue. To validate the proposal, the technique is simulated by using two different kinds of systems, as one is reserved for attacker and other one is for legitimate user. On the basis of simulation, we obtain very interesting statistical data, which show that MSAIDS improves the capturing performance and controls the attacks to be generated by DHCP rogue as compare with original IDS and other techniques. The findings demonstrate that MSAIDS has significantly reduced the false alarms. The findings demonstrate that MSAIDS has significantly reduced the false alarms. Finally, we analyze the overall efficiency of MSAIDS and existing technique by showing the results in tabular form and plotting the impact of results. In future, this technique will be deployed in measure the heart beats and other disease. Furthermore the broader impact of this research is to substantiate the MCL and improve the pedagogical activities and fostering other organizational requirements. This research will also boost the confidential level of the people while using MCL.

## BIOGRAPHY

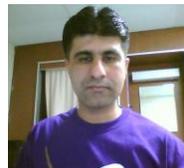

**Abdul Razaque** received Master degree in Computer Science from Mohammed Ali Jinnah University, Pakistan in 2008, and is a PhD student in of computer science and Engineering department in University of Bridgeport. His current research interests include the design and development of learning environment to support the learning about heterogamous domain, collaborative discovery learning and the development of mobile applications to support mobile collaborative learning (MCL), congestion mechanism of transmission of control protocol including various existing variants, delivery of multimedia applications. He has published over 50 research contributions in refereed conferences, international journals and books..

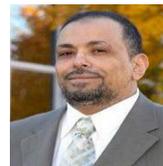

**Dr. Khaled Elleithy** is the Associate Dean for Graduate Studies in the School of Engineering at the University of Bridgeport. His research interests are in the areas of mobile wireless communications, mobile collaborative learning network security, formal approaches for design and verification. He has published more than two hundred research papers in international journals and conferences in his areas of expertise.. Dr. Elleithy is the editor or co-editor of 10 books published by Springer for advances on Innovations and Advanced Techniques in Systems, Computing Sciences and Software. He received the award of "Distinguished Professor of the Year", University of Bridgeport, during the academic year 2006-2007.

# APPENDIX:

Figure 12: Process of capering Anomaly based attacks with discrete simulation in C++

Figure 13: The process of capturing signature based attacks (IDS) with Discrete simulation in C++.